\documentclass[a4paper,10pt,twoside]{cpc-hepnp}
\usepackage{CJK,upgreek,fancyhdr}
\usepackage{multicol}
\usepackage{graphicx}
\usepackage{multirow}
\usepackage{booktabs}
\usepackage{amssymb,bm,mathrsfs,bbm,amscd}
\usepackage[tbtags]{amsmath}
\usepackage{lastpage}
\usepackage{color}

\begin{document}

\fancyhead[c]{\small Chinese Physics C~~~Vol. xx, No. x (201x) xxxxxx}
\fancyfoot[C]{\small 010xxx-\thepage}

\footnotetext[0]{Received  xx xx 201x}

\title{Simulation Study on the Emittance Compensation of Off-axis Emitted Beam in RF Photoinjector\thanks{Supported by the National Nature Science Foundation of China (11375199), and the Chinese Scholarship Council}}

\author{%
      Rui-Xuan Huang $^{1,2;1)}$\email{hruixuan@mail.ustc.edu.cn}%
\quad Chad Mitchell$^{2}$
\quad Qi-Ka Jia $^{1;2)}$\email{jiaqk@ustc.edu.cn}%
\quad Christos Papadopoulos$^{2}$
\quad Fernando Sannibale$^{2}$
}
\maketitle

\address{%
$^1$ NSRL, University of Science and Technology of China, Hefei, Anhui, 230029, China\\
$^2$ ATAP, Lawrence Berkeley National Laboratory, One Cyclotron Road, Berkeley, California, 94720, USA\\
}

\begin{abstract}

To make full use of photocathode material and improve its quantum efficiency lifetime, it can be necessary to operate laser away from the cathode center in photoinjectors. In RF guns, the off-axis emitted beam will see a time-dependent RF effect, which would generate a significant growth in transverse emittance. It has been demonstrated that such an emittance growth can be almost completely compensated by orienting the beam on a proper orbit in the downstream RF cavities along the injector\cite{PRAB paper}. In this paper we analyze in detail the simulation techniques used in reference\cite{PRAB paper} and the issues associated with them. The optimization of photoinjector systems involving off-axis beams is a challenging problem. To solve this problem, one needs advanced simulation tools including both genetic algorithms and an efficient algorithm for 3D space charge. In this paper, we report on simulation studies where the two codes ASTRA and IMPACT-T are used jointly to overcome these challenges, in order to optimize a system designed to compensate for the emittance growth in a beam emitted off axis. 

\end{abstract}

\begin{keyword}
photoinjector, off-axis emission, genetic algorithm, ASTRA, IMPACT-T, RF effect, emittance compensation
\end{keyword}

\begin{pacs}
29.27.-a,    02.60.Ej,   02.60.Pn
\end{pacs}

\footnotetext[0]{\hspace*{-3mm}\raisebox{0.3ex}{$\scriptstyle\copyright$}2013
Chinese Physical Society and the Institute of High Energy Physics
of the Chinese Academy of Sciences and the Institute
of Modern Physics of the Chinese Academy of Sciences and IOP Publishing Ltd}%

\begin{multicols}{2}

\section{Introduction}

	RF-gun-based photoinjectors are built to generate high brightness electron beams with low emittance and high charge, which can be used to drive Free Electron Lasers \cite{FEL1, FEL2, FEL3}, to generate THz radiation \cite{THz1, THz2, THz3}, and to probe structural dynamics at ultrafast time scale \cite{ultrafast}. Typically in photocathode guns, the laser excites electrons in the photocathode center area to get the best beam emittance performance. However, some situations, especially at high repetition rate or continuous wave operation, could require a beam emitted away from the cathode center. Primary reasons are quantum efficiency (QE) depletion in the cathode center that may develop after a number of hours of emission \cite{depletion}, and cathodes with a non-uniform QE distribution \cite{nonuniform}. Laser off-axis operation by making full use of the cathode area should significantly increase the QE lifetime and allow the cathode to operate for a much longer time.
For example, the Cornell DC photo-gun presently uses cathodes with an active area off-center to avoid damage due to ion back-bombardment 
\cite{Cornell}. When using off-axis emission in RF guns, the beam experiences a time dependent RF focusing that creates longitudinal-to-transverse correlations along the beam that ultimately generate a projected emittance increase.

An effective compensation mechanism for the emittance growth is reported in reference \cite{PRAB paper}, which states that the emittance growth caused by the time-dependent RF defocusing effect in the gun can be compensated by the time-dependent RF focusing force from the downstream RF cavities. In this paper, we describe the complex simulation set-up used in demonstrating the effectiveness of the compensation technique used in \cite{PRAB paper}. First, we present the multi-objective genetic algorithm (MOGA) that was used for optimizing the emittance compensation for the offset beam in the absence of space charge (SC). This MOGA procedure defines the proper setting for 4 couples of horizontal and vertical dipole correctors that place the beam on a particular orbit inside the downstream RF cavities. This is for the purpose of receiving a time-dependent RF focusing force which compensates for the one that the beam received inside the gun. Then, we discuss several simulation code issues, including the limitations of ASTRA's 2D space charge model and the limitations of IMPACT-T's dipole model for beams that are far off-axis. Finally in the last part of the paper, we describe the joint-ASTRA-IMPACTT combined simulation procedure developed to solve those issues, and to precisely optimize the emittance compensation of the off-axis emitted beam including space charge forces in the simulations.

\section{Off-axis beam emittance compensation by genetic algorithm}
\label{MOGA method}

In this section, we will introduce a multi-objective genetic algorithm as a useful tool to optimize high-brightness injector parameters. Then we optimize the performance of a beam emitted off-axis, using the APEX injector as an example. Space charge effects are not included in this section.

\subsection{Multi-objective genetic algorithm}

Genetic algorithms are inspired by characteristics in natural selection and heredity such as crossover and mutation \cite{GA}. The method of multi-objective genetic optimization (MOGA) is an effective approach to solve the problem with goals which are generally competing. The optimizer was written by integrating the genetic algorithm NSGA-II \cite{NSGA} together with the beam dynamics tool ASTRA \cite{ASTRA}, which could be applied to globally optimize high-brightness injector parameters. The optimizer is written primarily in the C language, and the program is integrated in Python which is able to call ASTRA simulation. MOGA has been actively used in APEX design with two objectives of minimizing the emittance and the bunch length. The optimizer typically runs in parallel on about 100 processors, once converged after adequate generations, it shows trade-offs between the emittance and bunch length.

The Advanced Photoinjector Experiment (APEX) \cite{APEX} at the Lawrence Berkeley National Laboratory is an injector R\&D facility aimed at testing the performance of a high brightness, high repetition rate VHF-gun \cite{VHFgun}. The APEX project is also the current baseline for LCLS-II injector \cite{LCLSii}. A 100~pC charge beam is emitted from the VHF-gun with proper emittance compensation in solenoids \cite{solenoid}, will be RF compressed by the buncher \cite{compress} and further accelerated through the TESLA-type RF cavities \cite{TESLA}. Finally at injector exit, the beam will be boosted to a 95~MeV energy with a sub-$\mu$m low emittance. A schematic layout of the beamline is plotted at \cite{PRAB paper}. And details in optimization procedure and setup for RF cavities, solenoids, etc. could be found elsewhere \cite{ChrisDesign}, which determines the nominal settings used in this paper (reference beam).

\subsection{Correction of emittance growth by MOGA}

Previously, we have introduced the genetic optimizer MOGA to obtain optimal settings in the APEX injector. Based on the nominal settings, the possible correction procedure for an off-axis beam is investigated to reduce the off-axis beam emittance growth. There are two pairs of dipole correctors located upstream of the buncher cavity and TESLA cavities respectively, which could be used to steer the beam trajectory into RF cavities. We will vary dipole settings to compensate the beam emittance induced by the RF effect, and show the correction results of optimization.

Assume that a beam with 2.0 mm misalignment in horizontal, ten times of the root mean square (rms) laser spot, is emitted on the cathode of the VHF-gun. Compared with a reference beam (on-axis emitted), the final emittance of the offset beam is increased by 3\% and 148\% in horizontal and vertical, respectively. Due to almost 90 degrees Larmor rotation \cite{Larmor} of two solenoids, the beam misalignment and emittance growth are exchanged in the two planes. Simulation shows the particle distribution at injector exit is transverse-longitudinal correlated. 

In order to obtain a minimum emittance solution at the injector exit, the bending radii (8 knobs) of the two pairs of correctors are adjusted to change the beam trajectory. The normalized rms horizontal emittance ($\epsilon_x$) and the vertical emittance ($\epsilon_y$) are chosen as two objectives to be minimized. As mentioned previously, the result is not a single solution, but instead a set of solutions with a trade-off between horizontal and vertical emittance. The final decision on the correction result is the one that gives the minimum value of the emittance geometric mean $\epsilon_G=\sqrt{\epsilon_x \epsilon_y}$. The minimum $\epsilon_G$ attained by optimization is 0.207~$\mu$m, while the corresponding values are 0.312~$\mu$m for the off-axis uncorrected beam and 0.196~$\mu$m for reference beam, respectively. It indicates that the optimization scheme could almost remove the emittance dilution due to off-axis emission. MOGA algorithm is a useful method to optimize dipole corrector setup for emittance compensation. More results and analysis are reported at \cite{PRAB paper}. 

Simulation shows that steering the beam back into axis alignment results in a larger transverse emittance. And optimized beam tracking shows the corrected beam maintains misalignment in the RF cavities. This means dipole corrector cannot directly reduce the emittance. It is the beam misalignment through the RF cavities that does the correction job. The beam emittance growth when laser off-axis emitted is due to time-dependent RF defocusing, which could be compensated by time-dependent RF focusing in other RF cavities. Since this RF effect is independent of SC effect, the correction procedure without consideration of space charge is still feasible. Further SC calculation will be included later.

\section{Simulation issues}

So far we have optimized the off-axis beam without SC calculation. However, the SC field may change the particle longitudinal distribution thus significantly influence the off-axis beam. ASTRA and IMPACT-T are both popular simulation tools to track the particles in an injector. During the simulation, we found the 2D space charge algorithm in ASTRA can not calculate the SC accurately when tracking the off-axis beam. While it is difficult for IMPACT-T to model dipole correctors. Before proceeding optimization with the SC effect, we will discuss specific restrictions on both of the codes and find the corresponding solutions.

\subsection{ASTRA limitation and its solution}

Since IMPACT-T has been demonstrated as an accurate beam dynamics code, it is a good benchmark for the ASTRA simulation results. ASTRA contains algorithms for both 2D (R-Z axisymmetric) and 3D space charge. For a reference beam with or without SC field, the 2D ASTRA simulation is precise and coincides with the 3D IMPACT-T result. But it is not true for the off-axis case. The Fig.~\ref{fig1} shows the transverse emittance comparison between the on-axis and off-axis beams, through SC simulations of the two codes. 

\begin{center}
\includegraphics[width=7.5cm]{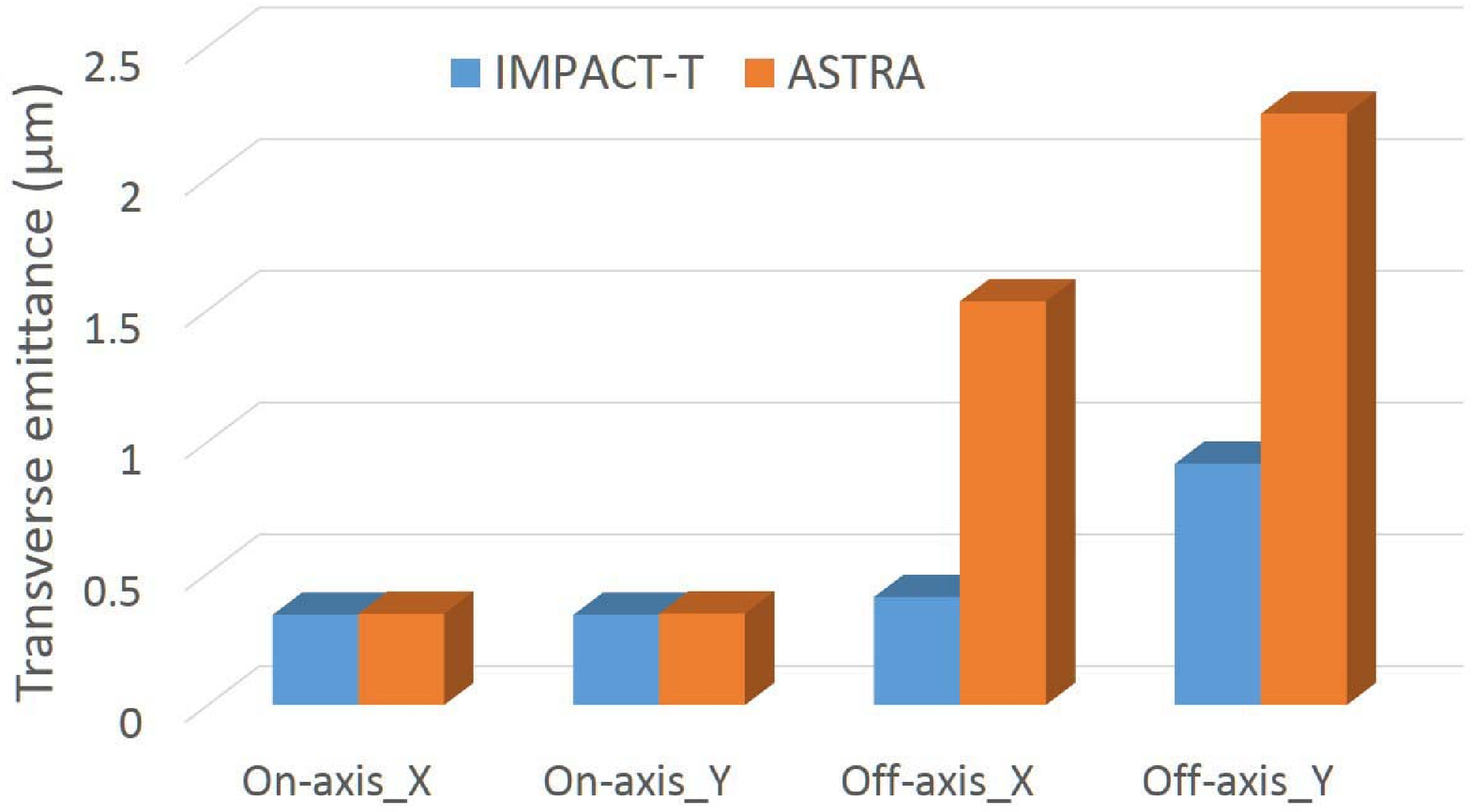}
\figcaption{(color online)  Comparison of space charge calculation by ASTRA and IMPACT-T}
\label{fig1}
\end{center}

For the on-axis case, the two codes give the results with a good agreement. While for the off-axis case, the 2D ASTRA and IMPACT-T show an obvious difference when including SC calculation. Because the 2D SC algorithm will introduce a spurious nonlinear SC field when dealing with a transverse-longitudinal correlated beam. Limitation sketch of 2D grid in the ASTRA SC calculation is shown in Fig.~\ref{fig2}. The cylindrical grid setup will assume a constant charge density inside a ring, which indicates the asymmetric beam would see a spurious nonlinear SC field. ASTRA will regard the maximum transverse dimension as the bunch size, and introduce an emittance calculation error. People could verify it by a simple simulation test.

\begin{center}
\includegraphics[width=6.5cm]{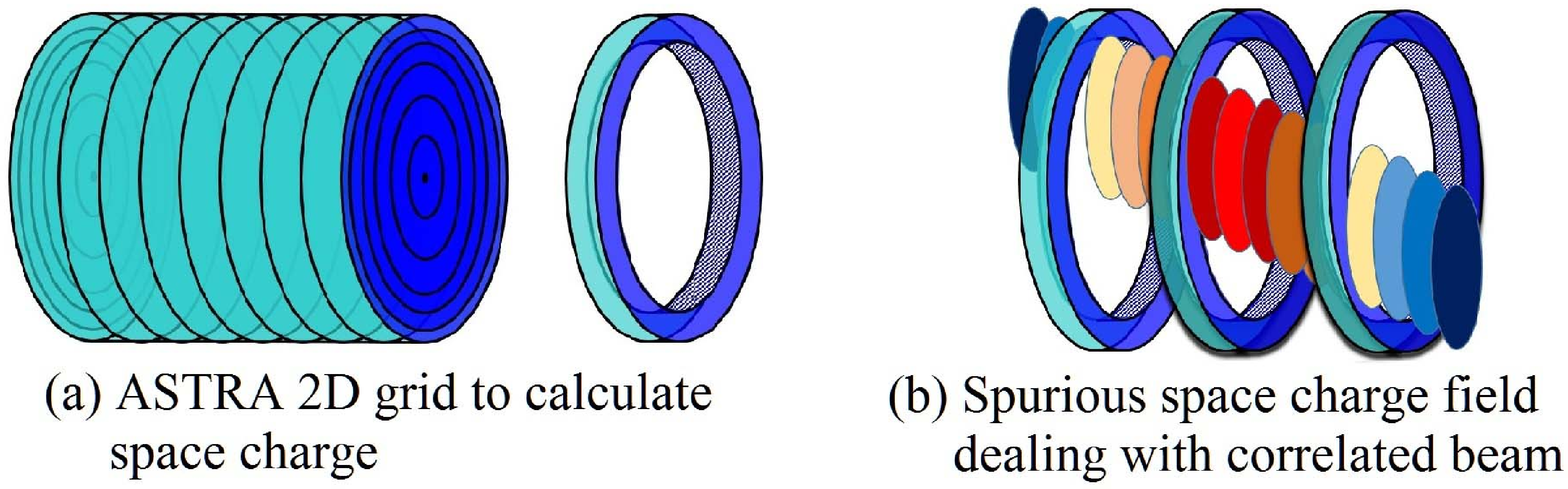}
\figcaption{(color online)  Schematic of ASTRA 2D cylindrical grid and the spurious space charge field dealing with the correlated beam}
\label{fig2}
\end{center}

There are two possible ways to solve the issue, they are

1) By ASTRA SC calculation with the 3D algorithm. ASTRA is capable of 3D SC calculation, with more grid numbers and more macroparticles, to get a sufficient statistical accuracy. However, the ASTRA 3D algorithm of the present version does not provide special features for the particle emission from a cathode, and the image charge force is not included. During the emission the grid setup has completed in a fairly short time. Overall, the 3D algorithm is restricted in SC calculation in the RF gun hence should not be used to simulate the off-axis beam.

2) By precise SC calculation with 3D IMPACT-T algorithm. IMPACT-T is a fast and accurate code using a 3D quasi-static model for high brightness beam dynamics simulation \cite{qiang2006three}. IMPACT-T can describe the off-axis beam precisely, especially for the case of beam emission from an RF gun. 

\subsection{IMPACT-T limitation and its solution}

We choose the IMPACT-T to simulate the off-axis beam due to its fast and accurate SC treatment but it's not that easy. There are also some limitations in the IMPACT-T optimization process. On the one hand, the MOGA optimizer is integrated with the 2D ASTRA, and we did not have access to a MOGA optimizer using IMPACT-T. On the other hand, there is not a straightforward way to add dipole correctors in IMPACT-T. In this subsection, we will discuss the restriction of the dipole setting and find a way to solve the problem.

The dipole element is included in IMPACT-T, and the bending magnet is modeled by an area of constant vertical magnetic field and two areas of fringe field on both dipole ends. One needs to define four linear equations as the pole faces, and several Enge parameters \cite{Enge} as the fringe region. Before the beam entering the magnet, a reference particle is defined by the beam centroid in the local coordinate system \cite{IMPACT}. Based on the initial longitudinal coordinates of the reference particle, both the macroparticles and the reference one are moving in the dipole magnetic field. When dealing with the SC calculation, all macroparticles coordinates are rotated into the coordinate system originating at the reference one. After finishing the bend, the macroparticles coordinates are transformed back to the coordinate system of the reference particle. However, the code only supports a horizontal bending ($B_y$) at the moment. People need to exchange $x$ and $y$ coordinates in advance if a vertical bending is required, and the code does not support repetitive coordinate exchanges.

Meanwhile, IMPACT-T assumes the reference particle will move through the axis of the dipole and into the axis of the next element \cite{IMPACT}, which means that the overall beamline will be modified. While the dipole corrector used in off-axis beam correction is to steer the beam trajectory and give the beam a certain offset into the next element. Anyhow, the dipole element is more suitable for a magnetic compressing or beamline bending instead of the beam correcting, thus should not be used to correct the off-axis beam. It could be certificated by a simple beam trajectory tracking.

To realize the beam correction, we propose to create the electromagnetic field distribution to represent a quasi-corrector. The element ``EMfldCart'' in IMPACT-T can read a discrete electromagnetic field ($E_x$, $E_y$, $E_z$, $B_x$, $B_y$, $B_z$) as a function of ($x$, $y$, $z$) from a ``T7'' or ``T8'' type file. IMPACT-T will solve the electron motion equations with contributions from both the external fields and the space charge fields, which presents the beam dynamic simulations in the global coordinate. After characterizing the field range, grid setting, peak field strength and geometric boundary, we could generate the required field data. 

\subsection{Dipole field characterization in two codes}

Since the MOGA has already optimized the setup for the dipole correctors, one would naturally think of imitating the dipole field in ASTRA and developing an equivalent electromagnetic field in IMPACT-T. It is necessary to compare the dipole field characteristics between the two codes. 

\subsubsection{The bending dipole setup in ASTRA}

A parameter ``Gap'' is defined to describe the dipole fringe field. The magnetic field in the transverse plane decays outside the dipole as \cite{ASTRA}:
\begin{equation}
B_{x,y}(d)=B_0 \left( 1+ \text{exp}^{4d/\text{Gap}}\right)^{-1},
\label{eq:Astra fringe field}
\end{equation}
where $B_0$ is the peak field of the dipole. And $d$ is the normal distance from the dipole edge, which has a maximum extension of 1.5 $\cdot \text{Gap}$.

We assume each corrector has a geometric length of 20~mm, and a ``Gap'' of 6.5~mm. The corresponding fringe field has an extension of 10~mm. Further increase of the ``Gap'' will bring about an ``overlap error'' and break the simulation down. People should make sure that the magnets in the two transverse planes do not overlap.

\subsubsection{The field data setup in IMPACT-T}

The magnet in IMPACT-T is also assumed as a length of 20~mm. When generating a magnetic field, the analytical formulas in \cite{FieldCreat} are referred to describe the fringe field. A ``gap'' parameter is provided as well (distinguished from the ``Gap''), unlike in ASTRA, the fringe field is extended at least 5 $\cdot \text{gap}$. A shorter extention of the fringe field will lead to an unreasonable emittance increase. For the same range of 10~mm fringe field and a negligible emittance increase, the gap is set as 0.5~mm with a high resolution of grid. The on-axis field distribution of the dipole corrector is shown in Fig.~\ref{fig3}.

\begin{center}
\includegraphics[width=6.5cm]{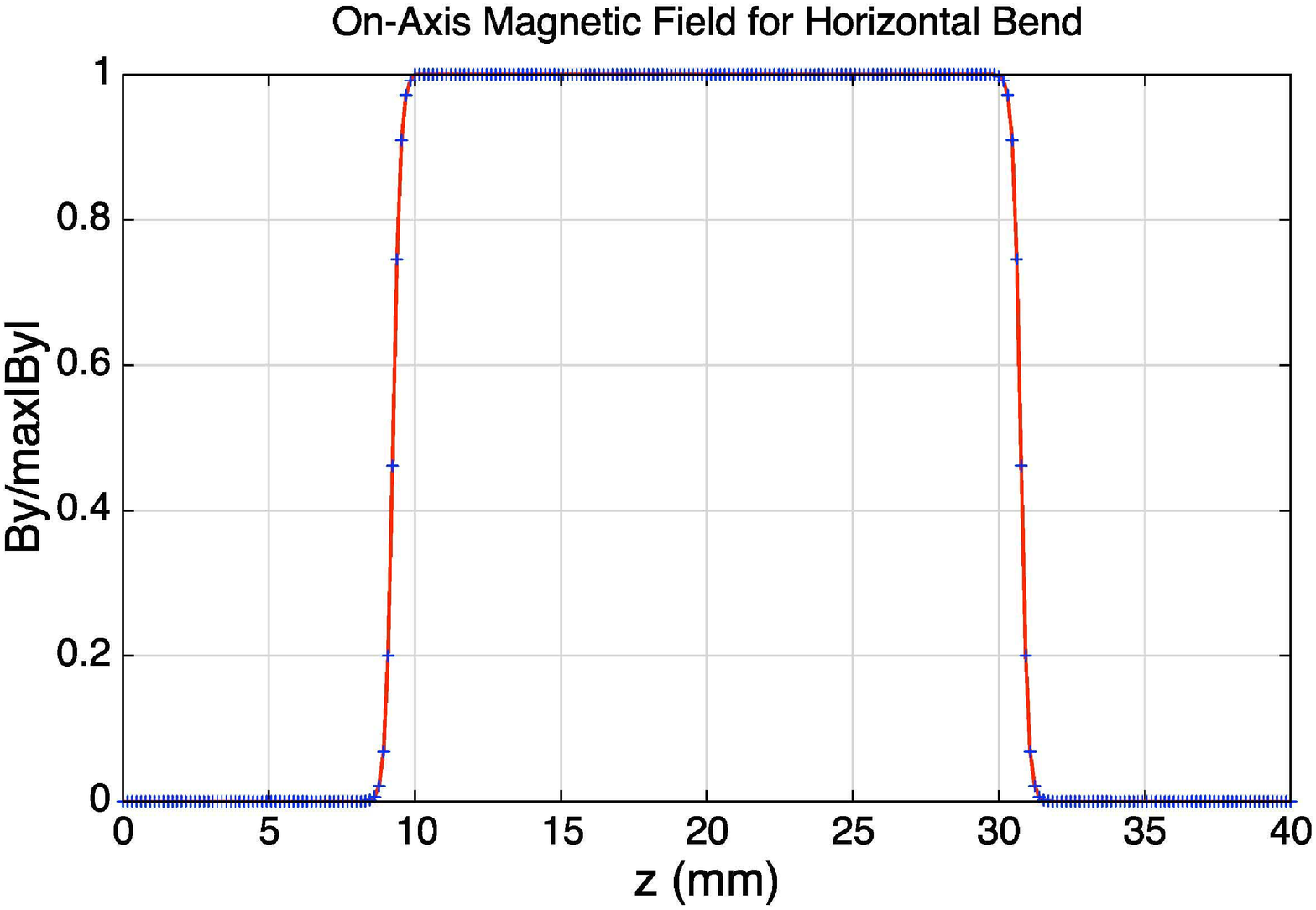}
\figcaption{(color online) IMPACT-T normalized magnetic field of the dipole corrector}
\label{fig3}
\end{center}

\section{The joint-ASTRA-IMPACTT optimization program}

In the previous section, we have discussed the limitations in the off-axis beam simulation using both ASTRA and IMPACT-T. The advantage of ASTRA is its integration with an available MOGA optimizer, which could select optimized solutions from tremendous results of beam dynamic simulations. While ASTRA can not calculate the SC precisely. The advantage of IMPACT-T is the accurate model for an efficient SC calculation of the off-axis beam. With the created magnetic field data, the correction procedure could also be realized. While we did not have access to a MOGA optimizer using IMPACT-T, and IMPACT-T is not able to scan multiple parameters simultaneously. Neither magnets optimization nor parameter scanning is available without the use of external scripts.

To combine their superiorities, and optimize the off-axis beam reliably, a joint-ASTRA-IMPACTT program is proposed. Before introducing the ``recipe'' of the joint program, the IMPACT-T simulation results should be benchmarked against the ASTRA ones.

\subsection{Codes benchmark}

We assume two identical beams go through a single dipole corrector in ASTRA and IMPACT-T respectively, and compare the evolution difference (without the SC calculation). In Fig.~\ref{fig4}, the transverse beam emittance and centroid evolutions are compared between ASTRA and IMPACT-T. Simulation shows that the beam emittance at the exit of the dipole is almost the same for the two codes, while the beam trajectory does not completely agree. Under the same magnetic strength, the IMPACT-T bends the beam more than the ASTRA does. It is manifested that one needs to decrease the peak field in IMPACT-T to benchmark against ASTRA. Figure \ref{fig4} provides some guidelines to adjust the field strength in IMPACT-T.

\begin{center}
\includegraphics[width=7.6cm]{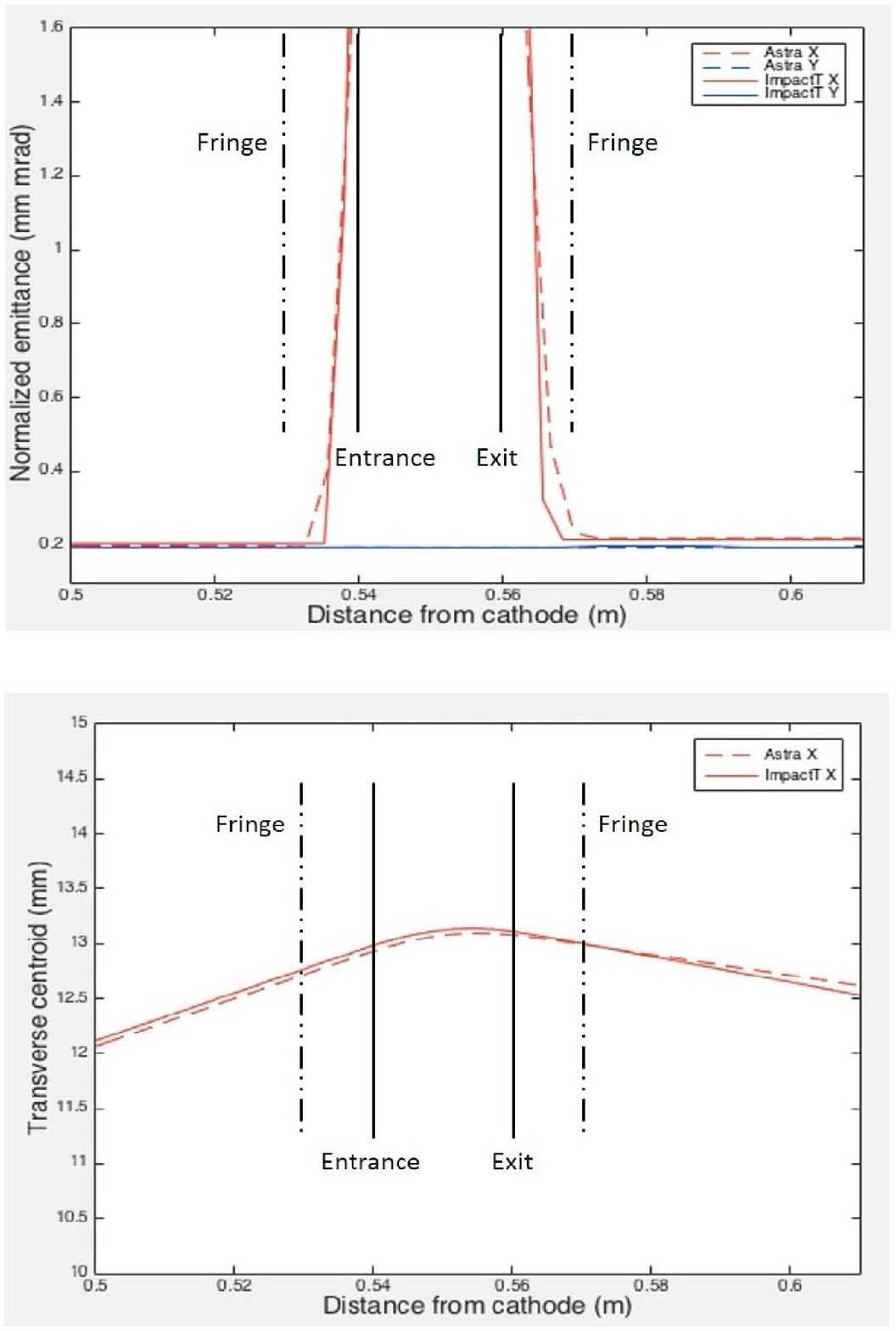}
\figcaption{(color online) Transverse emittance (top figure) and centroid  (bottom figure) evolutions comparison of ASTRA and IMPACT-T with a single dipole}
\label{fig4}
\end{center}

\subsection{Procedure of the joint simulation scheme}

The ``recipe'' of the joint simulation procedure is described as follows:

1) Tracking an off-axis beam (250k-particle) from the cathode to the gun exit, including a 3D SC calculation.

2) Randomly cutting down the particle number from 250k to 50k, transforming the distribution data from the IMPACT-T format to the ASTRA format.

3) Shutting down the SC, optimizing the off-axis beam by MOGA optimizer.

4) Based on the optimized bending radii of the correctors, tuning the field strengths in IMPACT-T carefully until the beam trajectories of the two codes are identical (without SC).

5) Once the magnetic settings in IMPACT-T are acquired, the 3D SC ``start-to-end'' off-axis beam simulation could be finally realized.

We firstly track the off-axis beam using IMPACT-T, because at very beginning in an RF gun, the beam has already been affected by the significant RF effect, thus a precise particle simulation with the 3D SC calculation is essential. The decrease of particle number in step 2) is conducive to the efficiency of the optimization. As the more particle number involved, the longer time that the optimizer will take to converge. A Matlab script is written to transform the distribution format automatically. For a complete beam tracking in the last step, 250k particle is suggested to evaluate the SC effect of the off-axis beam more accurately.

\subsection{Simulation results without space charge field}

According to the steps above, we will firstly follow the method in Section~{\ref{MOGA method}}, and update the MOGA optimization results using the initial particle distribution from IMPACT-T. Updated bending radii with tuned magnetic field strengths will be listed in Table~\ref{tab1}.

Based on the average beam energy, the bending radius is transformed to the magnetic field strength B as
\begin{equation}
B=\frac{\beta \gamma mc}{e R},
\label{eq:B R formula}
\end{equation}
where $c$ is speed of light, $m$ and $e$ are the electron rest mass and charge. $\beta$ is the ratio of the electron velocity to $c$, and $\gamma=1/\sqrt{1-\beta^2}$. Different position of the injector corresponds to the different beam energy thus a different  relationship between $B$ and $R$.

\begin{center}
\includegraphics[width=8.3cm]{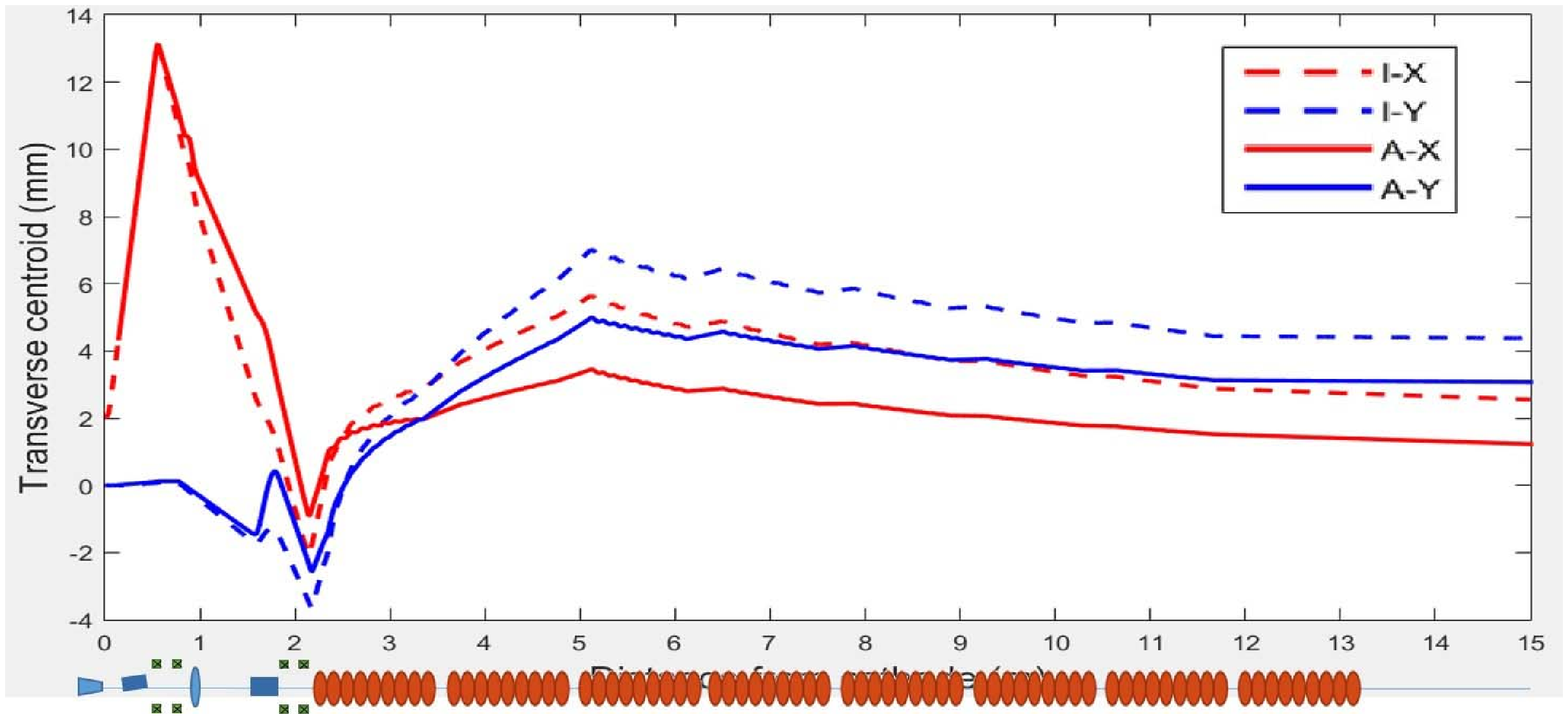}
\figcaption{(color online) IMPACT-T (dash lines) and ASTRA (solid lines) beam trajectory comparison with an identical magnetic field strength}
\label{fig5}
\end{center}

Secondly, by tracking the beam without SC in the two codes, we found that the relative difference of transverse emittance is 3.8\% and 23.1\% in the horizontal and vertical, respectively. It indicates that the beam emittance in the IMPACT-T simulation may not be fully compensated due to a possible trajectory misalignment in the dipoles. Further comparison of the beam transverse centroid evolutions is plotted in Fig.~\ref{fig5}. There is an obvious deviation in the beam trajectory of IMPACT-T. (The horizontal coordinate is the distance from the cathode. The location of each component could be found in the figure, and the corresponding representation will be explained in Fig.~\ref{fig6}.)

Thirdly, we need to finely adjust the magnetic strengths in IMPACT-T until the trajectories of two codes match. After the calibration, the beam centroid evolutions of the two codes are identical now. The optimized bending radii from MOGA, the magnetic strengths in ASTRA, and the magnetic strengths in IMPACT-T are listed in Table~\ref{tab1}. The first two correctors are located upstream of the buncher cavity, and the third and fourth correctors are located upstream of the first TESLA cavitiy. A negative bending radius represents a bending direction opposite to the positive one. Simulation shows that the trajectories of the two codes are identical after the IMPACT-T alignment, and the beam emittances at the injector exit are basically the same. It means that the emittance of the off-axis beam by IMPACT-T simulation is compensated as the MOGA expects. The final emittance of the offset beam is only increased by 3\% compared with the reference beam.

Overall, the joint-ASTRA-IMPACTT scheme is attainable and effective. By the MOGA optimization without consideration of space charge, the transverse emittance of the off-axis emitted beam is successfully compensated. If we compare the updated optimization result with the previous one in Section~{\ref{MOGA method}}, we found the beam emittance is further suppressed.

\end{multicols}
\begin{center}
\tabcaption{Correctors setting in MOGA, ASTRA, and IMPACT-T}
\label{tab1}
\footnotesize
\begin{tabular*}{140mm}{@{\extracolsep{\fill}}lccc}
\toprule Correctors & $R$ in MOGA /(m) & $B$ in ASTRA /(T) & $B$ in IMPACT-T /(T) \\
\hline
First corrector in horizontal & 0.643  & 0.00626  & 0.00587 \\
First corrector in vertical  & 99.876  & 4.03$\times 10^{-5}$  & 2.50$\times 10^{-5}$ \\
Second corrector in horizontal  & 14.889  & 2.71$\times 10^{-4}$  & 2.20$\times 10^{-4}$ \\
Second corrector in vertical  & 9.998  & 4.03$\times 10^{-4}$  & 3.70$\times 10^{-4}$ \\
Third corrector in horizontal  & -11.582  & 3.56$\times 10^{-4}$  & 3.57$\times 10^{-4}$ \\
Third corrector in vertical  & 51.141  & 8.06$\times 10^{-5}$  & 8.07$\times 10^{-5}$ \\
Fourth corrector in horizontal  & -0.996  & 0.00414  & 0.00382 \\
Fourth corrector in vertical  & -1.242  & 0.00332  & 0.00310 \\
\bottomrule
\end{tabular*}%
\end{center}

\begin{multicols}{2}

\end{multicols}

\vspace{0.1mm}

\ruleup
\begin{center}
\includegraphics[width=16.0cm]{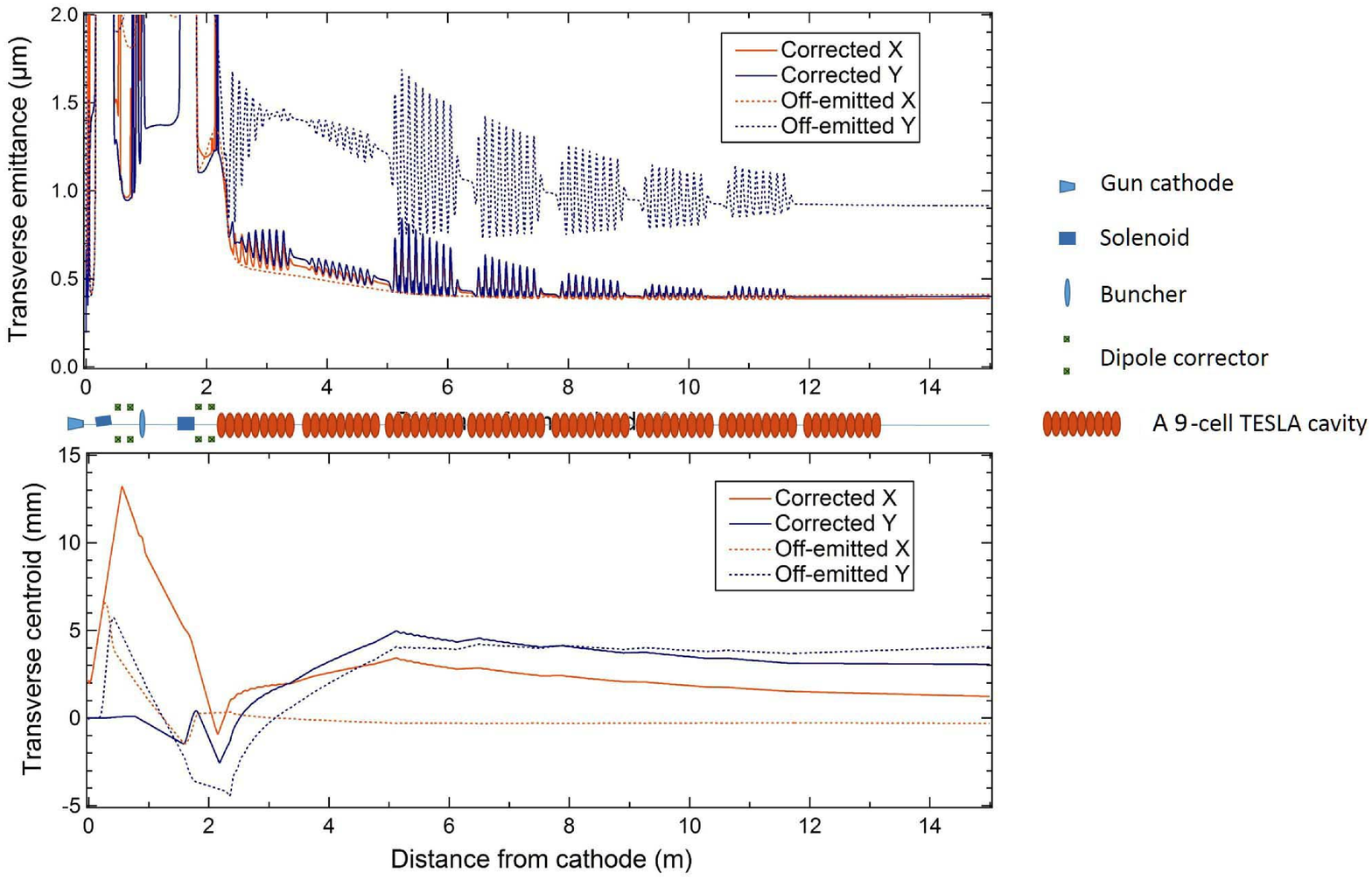}
\figcaption{(color online) Emittance (top figure) and centroid (bottom figure) evolution of optimized corrected beam, and the uncorrected beam. As the results of the joint-ASTRA-IMPACTT simulation program including 3D SC calculation.}
\label{fig6}
\end{center}
\ruledown

\vspace{0.1mm}

\begin{multicols}{2}

\subsection{Results of the joint simulation including  space charge effect}

It has been verified that the joint simulation program is operable, and the SC calculation of the off-axis beam should be finally available. 
With the same correctors setting as Table~\ref{tab1} and SC switched on, IMPACT-T will present the simulation in a large number of macroparticles. Firstly, the trajectory comparison between with and without SC shows that the transverse centroid evolution is not modified at all, which ensures the off-axis beam is optimized correctly by the dipole correctors.

In Fig.~\ref{fig6}, the rms emittances and the transverse beam centroids of the offset uncorrected beam and the offset optimized beam are compared. In the top figure, the transverse emittance of the optimized beam (solid lines) is remarkably reduced compared with the one before correction (dotted lines). It demonstrates that the joint optimization scheme could effectively compensate the emittance growth of the off-axis emitted beam. With SC consideration, the emittance growth (especially in the vertical plane) is greatly decreased from 168\% to 17\% (compared with the reference beam).

\section{Conclusion}

Research illustrated that a beam emitted off-axis from an RF photocathode gun experiences a time-dependent defocusing force, which leads to a transverse emittance increase \cite{PRAB paper}. The reference also describes a method to compensate for this emittance growth, by steering the beam on a proper orbit inside the other RF cavities downstream of the gun to impress an opposite time-dependent focusing effect. In this paper, we provide a detailed description of the simulation suite developed for the validation of the compensation procedure described above. A multi-objective genetic algorithm is utilized to optimally compensate for the transverse emittance increase in the beam emitted off-axis. We also describe the limitations associated with the two codes used in the study (ASTRA and IMPACT-T) and the simulation procedure that was developed to use the two codes in the proper sequence to avoid the limiting factors and provide the correct compensation results. 

\acknowledgments{The authors would like to thank Dr. Ji Qiang, Dr. Houjun Qian, Dr. Daniele Filippetto, Dr. Macro Venturini, and Dr. John Staples for insightful discussions.}

\end{multicols}

\vspace{-1mm}
\centerline{\rule{80mm}{0.1pt}}
\vspace{2mm}

\begin{multicols}{2}

\end{multicols}

\clearpage
\end{document}